\documentclass[aps,prb,twocolumn,showpacs,preprintnumbers]{revtex4-2}
\usepackage{graphicx}
\usepackage{amsmath}
\usepackage{amssymb}
\usepackage{graphicx,epstopdf}
\usepackage{graphicx}
\usepackage{epstopdf}
\usepackage{mathtools}
\usepackage{tikz}
\usepackage{color}
\usepackage[T1]{fontenc}
\usepackage{hyperref}
\usetikzlibrary{shapes.misc,shadows}

\newcommand{\be}{\begin{equation}}
\newcommand{\ee}{\end{equation}}
\newcommand{\bea}{\begin{eqnarray}}
\newcommand{\eea}{\end{eqnarray}}
\newcommand{\bse}{\begin{subequations}}
\newcommand{\ese}{\end{subequations}}

\begin{document}
\title {Steady state correlations and induced trapping of an inertial AOUP particle}

\author{N Arsha}
\author{K P Jepsin }
\author{M Sahoo}
\email{jolly.iopb@gmail.com}
\affiliation{Department of Physics, University of Kerala, Kariavattom, Thiruvananthapuram-$695581$, India}
\date{\today}
\begin{abstract}
We study the dynamics of an inertial active Ornstein-Uhnlenbeck particle self-propelling in a confined harmonic well. The transport behaviour of the particle is investigated by analyzing the particle trajectories, steady state correlations, and mean square displacement (MSD). The steady state correlation functions for the position as well as velocity are exactly calculated using different methods. We explore how the inertia affects the dynamical behaviour, when the particle is confined in a harmonic trap as well as when it is set free. From the exact calculation of MSD, it is observed that the initial time regimes are ballistic for both harmonically confined particle and free particle, whereas the long time regimes are diffusive for a free particle and non diffusive for a harmonically confined particle. One of our interesting observation is that the harmonically confined particle gets more and more confined with increase in the  self propulsion time or activity time of the dynamics and finally it gets trapped for very large self propulsion time. For a free particle, the velocity correlation decays by the complex interplay between the inertial time scale and the self propulsion time scale of the dynamics. Moreover, decorrelation in velocity happens only when these two time scales are of equal order.
\end{abstract}
\pacs{}
\maketitle
\section {\textbf{Introduction}}
In recent years, there is an immense interest in the study of systems which are driven far away from equilibrium. Active matter is one among these systems, which is inherently driven from equilibrium. The term 'active' refers to the development of directed motion generated by the constituents of the system by consuming energy from the local environment\cite{bechinger2016active,romanczuk2012active}. Examples of such active systems extend from motile cellular microorganisms like bacteria or unicellular protozoa\cite{boedeker2010quantitative}, artificially synthesized microswimmers like Janus particle \cite{walther2013janus}, micro robots \cite{palagi2018bioinspired}, hexbugs \cite{tapia2021trapped}, the flocking of birds\cite{cavagna2015flocking}, to school of fishes\cite{jhawar2020noise}. 
 One of the simplest and most extensively used standard model, namely active Brownian particle (ABP) model has been developed to treat the dynamical behaviour of particles in such active matter systems in both single particle level as well as collective level \cite{lobaskin2013collective,lowen2020inertial}. This term was first introduced in Ref.~\cite{schimansky1995structure}, where both the rotational as well as translational motion of the particles are taken into account. However, recently another model namely Active Ornstein-Uhlenbeck particle (AOUP) model \cite{caprini2020time,bonilla2019active,caprini2021inertial} is also used for studying the dynamical behaviour and steady state correlations of particles in such systems. The degree of irreversibility in the dynamical behaviour of an AOUP particle is also discussed in Ref. \cite{dabelow2021irreversibility}. This model was initially proposed by Uhlenbeck and Ornstein to study the velocity distribution of passive particles\cite{uhlenbeck1930theory}. The advantage of AOUP model is that it helps in the exact computation of analytical results \cite{sandford2018memory,sevilla2019generalized,singh2021crossover,bothe2021doi,caprini2021fluctuation,muhsin2021orbital}. Recently, the use of AOUP model has resulted many of the important features of active matters such as accumulation near boundary \cite{tian2017anomalous}, motility induced phase separation (MIPS) \cite{cates2015motility,speck2015dynamical,patch2017kinetics,wittkowski2014scalar} etc. 
 However, in most of the studies, inertia is not taken into account.
 Systems which are macroscopic in nature, while self-propelling in a dilute/gaseous or lower viscous medium, the inertia becomes prominent and plays an important role in the dynamics. The inertial influence in such systems challenge the theoretical modelling of the dynamical behaviour. Macroscopic swimmers \cite{leoni2020surfing,gazzola2014scaling,saadat2017rules,shahsavan2020bioinspired}, flying insects \cite{sane2003aerodynamics} etc. are some examples of this category, where inertia plays an important role. Hence the introduction of inertia to the standard models is necessary for treating the dynamics of such systems. Recently, inclusion of inertia in the ABP model results some modification in the fundamental properties of active systems, such as inertial delay \cite{scholz2018inertial}, motility induced phase transitions \cite{su2021inertia} and so on. So far the existing results on active matter systems are mostly on the over damped regime of the particle dynamics. The inertial effects of active matter systems have received only limited attention to the steady state correlation studies. Without the knowledge of inertia and inertial effects, the exact transport properties of a system will remain unexplored. Hence in this work, we are interested the dynamics of an inertial active Ornstein-Uhlenbeck particle moving in a viscous medium. We are mainly interested in steady state correlation functions and the mean square displacement of the dynamics. The exact computation of steady state correlation functions can provide the complete information about the dynamical behaviour or transport properties of the system. The analytical behaviour of steady state correlation functions of an overdamped active Brownian particle as well as AOUP particle are discussed in \cite{szamel2014self,sevilla2019generalized}. In this paper, we consider the inertia in the AOUP model and exactly compute the steady state correlation functions using various methods which is discussed in the following sections.
 \section{\textbf{Model}}
We consider the motion of an inertial active Ornstein- Uhlenbeck particle confined in an one dimensional harmonic trap, $U(x,t)$. The dynamics of the particle can be described by the Langevin equation of motion \cite{tothova2011langevin,noushad2021velocity,jayannavar2007charged,muhsin2021orbital,muhsin2022inertial}, which is given by
\begin{equation}
m\frac{dv}{dt}=-\gamma v-\frac{\partial U\left( x,t\right) }{\partial x}+\sqrt{D} \eta\left(t\right).\label{dynamics}
\end{equation}
Here, $m$ and $v(t)$ represent the mass and instantaneous velocity of the particle. $\gamma$ is the viscous coefficient of the medium. $\eta(t)$ is the  random noise which follows the Ornstein- Uhlenbeck process and satisfies the properties $<\eta(t)>=0$ and $<\eta(t)\eta(t')>=\dfrac{1}{2t_{c}}e^{-\frac{\vert t-t^{'}\vert}{t_{c}}}$, with $t_{c}$ being the noise correlation time or activity time in the dynamics. It is the time up to which there exists a finite correlation or activity in the dynamics. As a result the particle persists to self propel upto time $t_{c}$. A finite $t_{c}$ especially quantifies the activity or correlation in the medium, which decays exponentially with $t_{c}$. The angular brackets $\left\langle ...\right\rangle$ represent the ensemble average over the noise. $D$ is the parameter which remains constant through out the dynamics and represents strength of the Ornstein-Uhnelbeck noise \cite{nguyen2021active}. For a finite $t_{c}$, the system is in non-equilibrium and can't be mapped to the thermal equilibrium limit of the dynamics. However, in the white noise limit, i.e., $t_{c} \rightarrow 0$ limit, we consider $D=2\gamma k_{B}T$, with $T$ as the temperature of the system, to have the thermal equilibrium limit of the dynamics. The potential $U(x,t)(=\frac{1}{2} k x^{2})$ represents the harmonic potential, with $k$ as the harmonic stiffness constant. Substituting $k=m\omega_{0}^{2}$ or $U(x,t)=\frac{1}{2} m\omega_{0}^{2}x^{2}$ with $\omega_{0}$ as the harmonic frequency, the dynamics [Eq.~\eqref{dynamics}] can take the form as\\
\begin{equation}
\begin{split}
 m\frac{dv}{dt}=-m\omega_{0}^{2}x-\gamma v+\sqrt{D} \eta\left( t\right). \label{model1}
            \end{split}
         \end{equation}
\section{\textbf{RESULTS AND DISCUSSION}}
\begin{figure}[htp]
\centering
\includegraphics[scale=0.28]{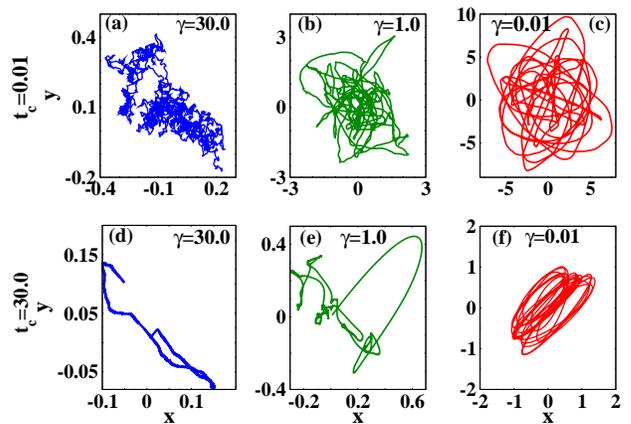}
\caption{ Simulated two-dimensional(2D) trajectories of a single particle moving in a harmonic confinement. (a), (b) and (c) are for $t_{c}=0.01$ and (d),(e) and (f) are for $t_{c}=30$. The other common parameters are $m=D=\omega_{0}=1$}
\label{trajectory harmonic}
\end{figure}
 By using the Fourier transform method formalism, the solution of the dynamics[Eq.~\eqref{model1}] in Fourier space $x(\omega)\left[=\int_{-\infty}^{\infty}e^{i\omega t}x(t)dt\right]$ and $v(\omega)\left[=\int_{-\infty}^{\infty}e^{i\omega t}v(t)dt\right]$ can be obtained as 
 \begin{eqnarray}\nonumber
x\left( \omega \right) =\dfrac{\sqrt{D}\eta \left( \omega \right)}{-m\omega^{2}+i\gamma \omega+\dfrac{\omega_{0}^{2}}{m}}
\end{eqnarray}
and\\
\begin{eqnarray}\nonumber
v\left( \omega \right) = \dfrac{\omega\sqrt{D}\eta \left( \omega \right)}{im\omega^{2}+\gamma \omega-i\omega_{0}^{2}m},
\end{eqnarray}
respectively with $\eta(\omega)\left[=\int_{-\infty}^{\infty}e^{i\omega t}\eta(t)dt\right]$. The time evolution of the dynamics is mainly controlled by the complex interplay of inertial force, self propulsion force and the force exerted by the harmonic confinement. In Fig.~\ref{trajectory harmonic} we show the simulated trajectories of the motion [Eq. (2)] of a single particle in xy-plane for different values of $\gamma$ and for different $t_{c}$ values. When $t_{c}$ is very small, the particle is mostly influenced by the inertial force and the harmonic confinement. Hence, for small $t_{c}$, when $\gamma$ is large, the inertial influence is negligibly small and the particle behaves like an over damped passive particle showing random trajectories. However, with decrease in $\gamma$ value, the inertial influence becomes prominent and the particle shows almost circular trajectories in the harmonic confinement. It changes it's direction slowly as compared to the higher $\gamma$ regimes resulting the trajectories comparatively smoother. In the inertial regime (i.e., for low $\gamma$ value), with increase in $t_{c}$ value, the particle gets more and more confined. As a result for larger $t_{c}$, the circular trajectories get squeezed.
Next we are interested in exploring the steady state position correlation and velocity correlation. The spectral density corresponding to position is given by the relation $S_{x}(\omega)[=\left\langle  x(\omega)x^{*}(\omega)\right\rangle]$, which is calculated as
\begin{equation}
    S_{x}(\omega)=\frac{D\left\langle \eta(\omega)\eta^{*}(\omega)\right\rangle}{m^{2}\left(\omega_{0}^{2}-\omega^{2}\right)^{2}+\omega^{2}\gamma^{2}},\label{sxw1}
 \end{equation}
 where
$$\left\langle \eta(\omega)\eta^{*}(\omega)\right\rangle=\int_{-\infty}^{\infty}e^{i\omega t}\left\langle \eta(t)\eta(0)\right\rangle dt.$$
Substituting the integral value as $\left\langle  \eta(\omega)\eta^{*}(\omega)\right\rangle= \frac{1}{1+\omega^{2}t_{c}^{2}}$ and simplifying Eq.~\eqref{sxw1}, $S_{x}(\omega)$ can be expressed as
\begin{equation}
S_{x}\left( \omega\right) =\dfrac{D}{(1+\omega^{2}t_{c}^{2})\left\lbrace (m \omega_{0}^{2}-m \omega^{2})^{2}+\omega^2\gamma^2\right\rbrace }. \label{sxw2}
\end{equation}

\begin{figure}[htp]
\centering
\includegraphics[scale=0.65]{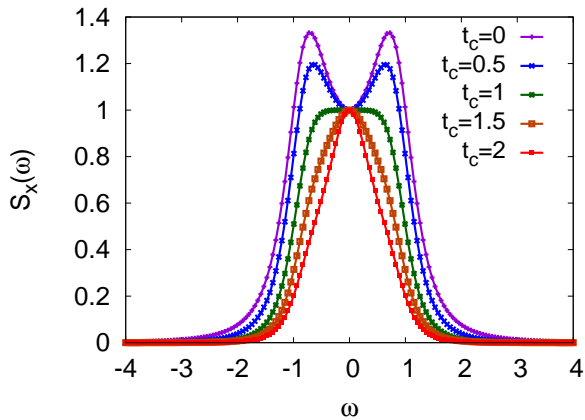}
\caption{ $S_x(\omega)$ [Eq.~\eqref{sxw2}] as a function of $\omega$ for different values of $t_{c}$. The other fixed common parameters are $D=1$, $m=1$, and $\gamma =1$.}
\label{spectral density position}
\end{figure}
In Fig.~\ref{spectral density position}, we have plotted the spectrum corresponding to the position $S_{x}(\omega)$ as a function of $\omega$ for various values $t_{c}$. For a particular value of $t_{c}$, the spectrum shows a non monotonic behaviour with $\omega$. For lower $t_{c}$ values, $S_{x}(\omega)$ shows a double peaking behaviour. Subsequently with increase in $t_{c}$ value, the double peaking behaviour gets suppressed and the spectrum becomes single peaked structure across $\omega=0$. With further increase in $t_{c}$, the peak gets sharper across $\omega=0$ as expected. This might be due to the reason that the particle gets more and more confined in the well with increase in $t_{c}$ value. The same conclusion is drawn from the particle trajectory plot in Fig.~1 for larger $t_{c}$ value. When the correlation time approaches zero, the active fluctuations become thermal and the particle behaves like a passive particle. In this limit, the analytical expression for $S_{x}(\omega)$ is consistence with that reported in Ref.\cite{noushad2021velocity} in the absence of any external force and the spectrum shows the same behavior as reported in Ref. \cite{noushad2021velocity}. The Optimum value of $\omega$ ($\omega_{m}$) at which the spectrum shows peaks can be obtained by taking the time derivative of $S_{x}(\omega)$ with respect to $\omega$ and equating it to zero, i.e., $\frac{dS_{x}(\omega)}{d\omega}\vert_{\omega=\omega_{m}}=0$. By simplifying this, it is confirmed that the optimum $\omega$ depends on the parameters $t_{c}$, $\omega_{0}$ and $\gamma$ values. With increase in $t_{c}$, $\omega_{m}$ decreases as a result the peak shifts towards $\omega=0$ and becomes single peaked at $\omega=0$. Similarly the  spectral density corresponds to velocity $S_{v}(\omega)\left[=\left\langle  v(\omega)v^{*}(\omega)\right\rangle\right]$ can be obtained as
\begin{equation}
S_{v}\left( \omega\right) =\dfrac{D\omega^{2}}{(1+\omega^{2}t_{c}^{2})\left\lbrace (m \omega_{0}^{2}-m \omega^{2})^{2}+\omega^2\gamma^2\right\rbrace }\label{svw}
\end{equation}
\begin{figure}[htp!]
\centering
\includegraphics[scale=0.65]{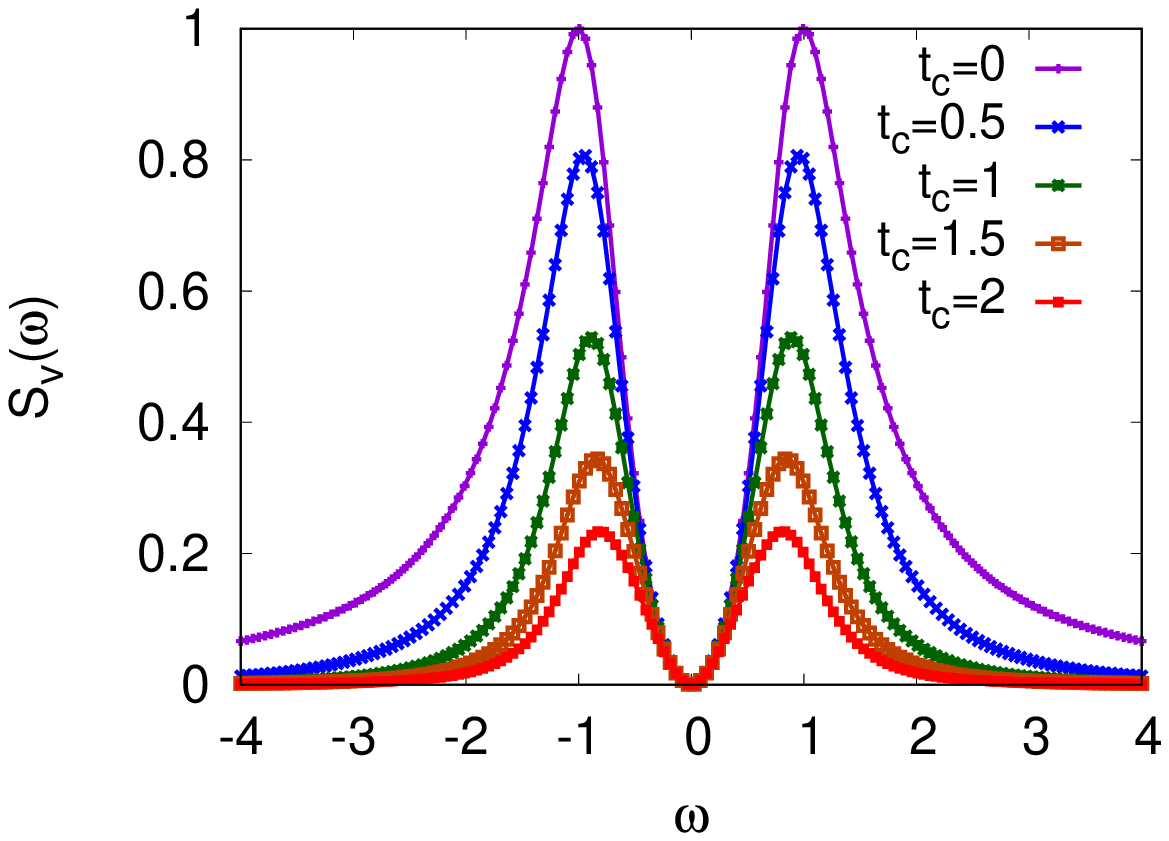}
\caption{ $S_v(\omega)$ [Eq.~\eqref{svw}] as a function of $\omega$ for different values of $t_{c}$. The other fixed parameters are $D=1, m=1\text{ and }\gamma =1$}
\label{spectral density velocity}
\end{figure}
In Fig.~\ref{spectral density velocity}, we have shown the plot for $S_{v}(\omega)$ as a function of $\omega$ for different values of $t_{c}$. For a given $t_{c}$ value, the spectrum is symmetric across $\omega=0$ and shows a double peaking behaviour with exponential tails. With further increase in $t_{c}$, the peaks get suppressed as expected. At the same time, the peaks get narrow and shifts towards lower values of $\omega$. For very large $t_{c}$ value, the influence of the harmonic confinement becomes stronger. As a result the exponential tails also get stiffer. This reflects the stronger effect of the barriers with longer persistence of activity in the dynamics. Therefore, the particle gets more and more confined with $t_{c}$. The optimum value of $\omega$ ($\omega_{n}$)at which the spectrum shows maximum can be obtained by simplifying $\dfrac{dS_{v}(\omega)}{d\omega}\vert_{\omega=\omega_{n}}=0$. 
It is confirmed that the optimum $\omega$ ($\omega_{n}$) at which the spectrum shows maximum decreases with $t_{c}$. Other than $t_{c}$, it also depend on the parameters $\omega_{0}$ and $\gamma$. 

The position correlation $C_{x}(t)$ and velocity correlation $C_{v}(t)$ can be obtained  by performing the inverse Fourier transform of $S_{x}({\omega})$ [Eq.~\eqref{sxw2}] and $S_{v}(\omega)$ [Eq.~\eqref{svw}] as $\int_{-\infty}^{\infty}e^{-i\omega t}S_{x}(\omega)d \omega$ and$ \int_{-\infty}^{\infty}e^{-i\omega t}S_{v}(\omega)d \omega$, respectively. Simplifying the integral $\int_{-\infty}^{\infty}e^{-i\omega t}S_{x}(\omega)d \omega$, $C_{x}(t)$ can be calculated as   
    \begin{equation}
    C_{x}(t)=\dfrac{t_{c}^{2}f_{1}}{2} e^{\dfrac{-t}{t_{c}}}+\dfrac{N}{4\omega_{0}^{2}}e^{\dfrac{-t}{2t_{m}}}\left(f_{2}\cos(\omega_{1}t)+f_{3}\sin(\omega_{1}t)\right),\label{cxtharmonic}
\end{equation}
where
\begin{widetext}
\begin{equation*}
\begin{split}
f_{1}=\dfrac{Dt_{c}}{(m+mt_{c}^{2}\omega_{0}^{2})^{2}-\gamma^{2}t_{c}^{2})} \text{, }  
N=\dfrac{D}{m\gamma \omega_{1}\left[1+t_{c}^{2}\left(2 \omega_{0}^{2}-\dfrac{1}{t_{m}^{2}}+t_{c}^{2} \omega_{0}^{4} \right) \right] }\text{, } \\
 f_{2} =\left[\omega_{1}\left( 2+2 \omega_{1}^{2}t_{c}^{2}-\dfrac{3t_{c}^{2}}{2t_{m}^{2}}\right)  \right] \text{ and }
f_{3}=\left[\dfrac{1}{t_{m}}+t_{c}^{2} \left( \dfrac{3\omega_{1}^{2}}{t_{m}}-\dfrac{\gamma}{4t_{m}^{2}}\right) \right].
\end{split}
\end{equation*}
\end{widetext}
Similarly simplifying the integral $ \int_{-\infty}^{\infty}e^{-i\omega t}S_{v}(\omega)d \omega$, the velocity correlation $C_{v}(t)$  can also be obtained as\\
\begin{equation}
C_{v}(t)=\dfrac{-f_{1}}{2m^{2}}e^{\frac{-t}{t_{c}}}+\frac{N}{4}e^{\frac{- t}{2t_{m}}}\left( f_{4} \cos(\omega_{1}t)+f_{5}\sin(\omega_{1}t)\right),\label{cvtharmonic}
\end{equation}
with
\begin{widetext}
\begin{equation*}
f_{4}=\left[\omega_{1}\left(2+2t_{c}^{2}\omega_{1}^{2}+\dfrac{t_{c}^{2}}{2t_{m}^{2}} \right)\right]  \text{ and } 
f_{5}=\left[-\dfrac{1}{t_{m}}+  t_{c}^{2}\left(\dfrac{\omega_{1}^{2}}{t_{m}}+\dfrac{1}{4t_{m}^{3}}\right)\right].
\end{equation*}
\end{widetext}
In both Eqs.~\eqref{cxtharmonic} and \eqref{cvtharmonic}, $\omega_{1}=\sqrt{\omega_{0}^{2}-\frac{1}{4t_{m}^{2}}}$ and $t_{m}=\frac{m}{\gamma}$. Using the correlation matrix formalism method, we have also obtained the same results for both $C_{x}(t)$ and $C_{v}(t)$, which is shown in appendix-A. The results for both $C_{x}(t)$ and $C_{v}(t)$ are well agreement with the athermal part of the correlations as reported in Ref. \cite{caprini2021inertial}
\begin{figure*}[hbtp]
\centering
\includegraphics[scale=1.]{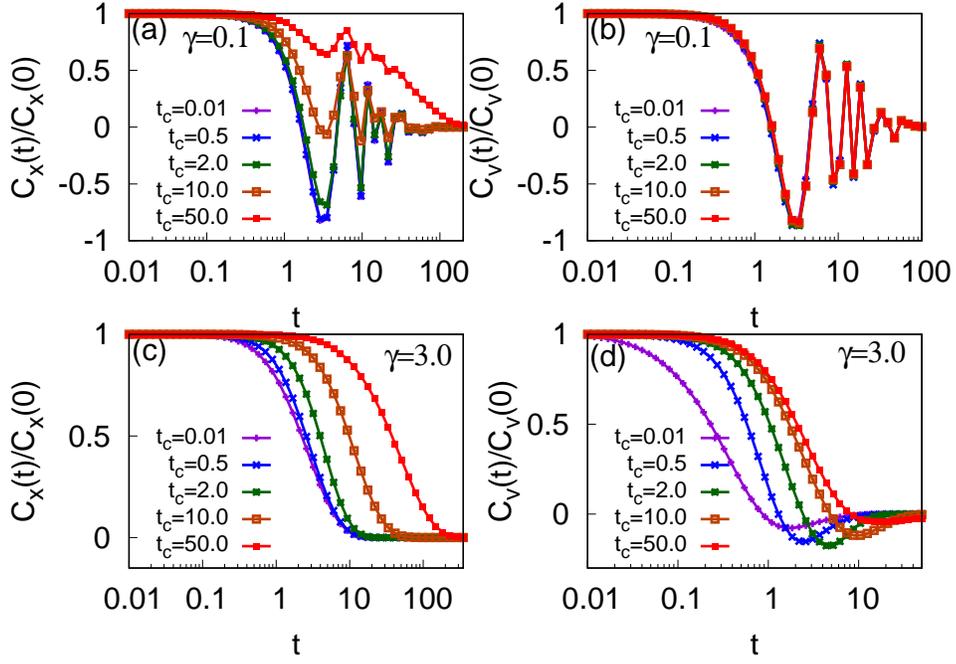}
\caption {Normalized $C_{x}(t)[\text{Eq}.~\eqref{cxtharmonic}]$ as a function of $t$ in (a) and (c) and the normalized $C_{v}(t)[\text{Eq}.~\eqref{cvtharmonic}]$ as a function of $t$ in (b) and (d) are shown for a free particle and for a confined harmonic particle, respectively for different $t_{c}$ values. The common parameters are $D=\omega_{0}=m =1$.}
\label{position and velo corr harmonic}
\end{figure*}
In Fig.~\ref{position and velo corr harmonic}, both normalized $C_{x}(t)$ and $C_{v}(t)$ are plotted as a function of $t$ for lower values of $\gamma$ in (a) and (b) and for larger values of $\gamma$ in (c) and (d), respectively. For low viscous medium ($\gamma=0.1$), where the inertial influence is prominent, both $C_{x}(t)$ and $C_{v}(t)$ persists with a finite value for initial time regimes and show an oscillatory behaviour in the long time regime of the dynamics before approaching to zero. For $C_{x}(t)$, the amplitude of the oscillation decreases with increase in $t_{c}$ values and the curves shift in such a way that for very large $t_{c}$, the correlation never approaches negative values (see Fig.~4(a)). However, for $C_{v}(t)$, the amplitude of the oscillation decreases with time. At the same time, the minimum value shifts towards right and approaches zero in the long time limit, which can be reflected from Fig.~ 4(b). For the case of a highly viscous medium (say for $\gamma=3.0$), where the inertial influence is negligible, the $C_{x}(t)$ never becomes negative and show a finite constant value before decaying to zero. With increase in $t_{c}$, the correlation persists for a longer time before decaying to zero (as in Fig.~4(c)). On the other hand, for very low $t_{c}$ value, $C_{v}(t)$ shows a damped oscillatory behaviour. It decays faster, becomes negative, attains a minimum value and then approaches zero. With further increase in $t_{c}$, the correlation persists for longer time and the decay becomes slower. At the same time, the minimum value shifts towards right and finally approaches zero for large $t_{c}$ (see Fig.~4(d)). In $t_{c}\rightarrow{0}$ limit, the result for $C_{v}(t)$ is consistent with that reported in Ref.~\onlinecite{noushad2021velocity} in the absence of any external force.\\
When the particle is not confined by the harmonic potential, it can freely self propel in the environment. The dynamics of the particle [Eq.~\eqref{model1}] in such a case can be reduced as,
   \begin{equation}
       m\frac{dv}{dt}=-\gamma v+\sqrt{D} \eta\left(t\right),\\\label{model2},
        \end{equation}
       \begin{equation}
            t_{c} \dot \eta(t)=-\eta(t)+\zeta(t).\label{noise}
        \end{equation}
        \begin{figure}[htp]
\centering
\includegraphics[scale=0.30]{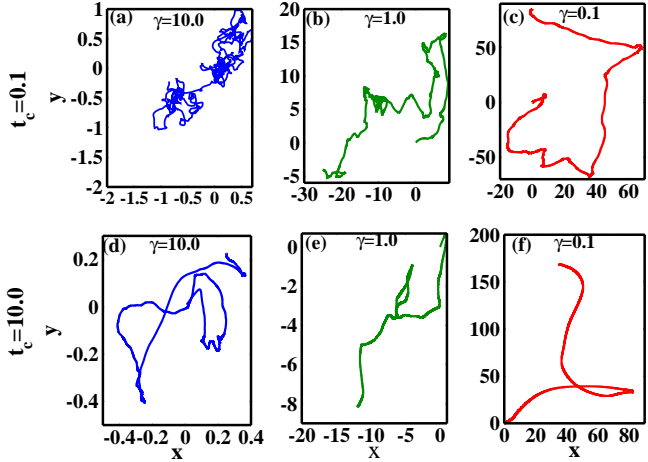}
\caption{2D single particle trajectories of a free particle. $t_{c}=0.1$ is considered for (a),(b) and (c) and $t_{c}=10$ is for (d),(e) and (f), respectively. The other common parameters are $m=D=1$}
\label{trajectoryfree}
\end{figure}\\
      The dynamics [Eq.~\eqref{model2}] and [Eq.~\eqref{noise}] can be written in the form \cite{caprini2021inertial},
        \begin{equation}
            \dot{w}=-A\cdot w+\sqrt{2} \sigma \cdot \eta,\label{matrix form}
        \end{equation}
        where,
    \[
     w=\left( \begin{array}{c}
       v\\
       \eta
       \end{array} \right),
        A=\left( \begin{array}{cc}
        
\frac{\gamma}{m} & \frac{-\sqrt{D}}{m} \\
0& \frac{1}{t_{c}} 
\end{array} \right), \text{and }
\sigma=
\left( \begin{array}{cc}
0& 0 \\
0& \frac{1}{\sqrt{2}t_{c}} 
\end{array} \right)
\]
With the help of correlation matrix method formalism\cite{caprini2021inertial}, by introducing the correlation matrix $C$, Eq.~\eqref{matrix form} can take the form as
 \begin{equation}
    A \cdot C+C\cdot A^{T}=2M,\label{correlation matrix form}
\end{equation}
where  $M=\sigma \cdot \sigma^{T} $ is the diffusion matrix and 
\[
C=
\begin{pmatrix}
C_{vv} & C_{v\eta}\\
C_{\eta v} & C_{\eta \eta}
\end{pmatrix}.
\]
Solving Eq.~\eqref{correlation matrix form}, the elements of the correlation matrix can be obtained as
    $C_{vv}=\frac{D}{2\gamma^{2}(t_{m}+t_{c})}$, $C_{v\eta}=C_{\eta v}=\frac{\sqrt{D}}{2\gamma (t_{m}+t_{c})}$, and $C_{\eta \eta}=\frac{1}{2t_{c}}$ respectively.
 The time dependent auto correlation function for velocity can be obtained from Eq.~\eqref{model2} by multiplying initial velocity $v_{0}$ to both the sides of Eq.~\eqref{model2}and taking the time derivative of average of each terms as,  
      \begin{equation}
          \frac{d}{dt}\langle v(t)v(0) \rangle+\frac{\gamma}{m}\langle v(t)v(0)\rangle=\frac{\sqrt{D}}{m} \langle \eta (t)v(0)\rangle.\label{derivative velocitycorre}
      \end{equation}
  Similarly, from Eq.~\eqref{noise} the time derivative of $\langle \eta(t) v(0)\rangle$ can be written as
      \begin{equation}
          \frac{d}{dt}\langle \eta (t)v(0)\rangle=-\frac{1}{t_{c}}\langle \eta (t)v(0)\rangle.\label{derivative noisecorre}
      \end{equation}
      By solving the above equation, we get
      \begin{eqnarray}
      \langle \eta (t)v(0) \rangle=C_{\eta v} \exp{\frac{-t}{t_{c}}}.
      \end{eqnarray}
     Now substituting the value for $C_{\eta v}$ in Eq.~\eqref{derivative noisecorre} and using the expression of $\langle \eta (t)v(0) \rangle$ in Eq.~\eqref{derivative velocitycorre}, one can solve Eq.~\eqref{derivative velocitycorre}for the velocity correlation $C_{v}(t)$($\langle v(t)v(0) \rangle$) as
\begin{equation}
   C_{v}(t)= \dfrac{D}{2\gamma^2 \left(t_{m}^{2}-t_{c}^{2}\right)}\left[t_{m}e^{-\dfrac{t}{t_{m}}}-t_{c}e^{-\dfrac{t}{t_{c}}}\right].\label{velocity correlation free}
\end{equation}
\begin{figure*}[ht!]
    \centering
    \includegraphics[scale=1.]{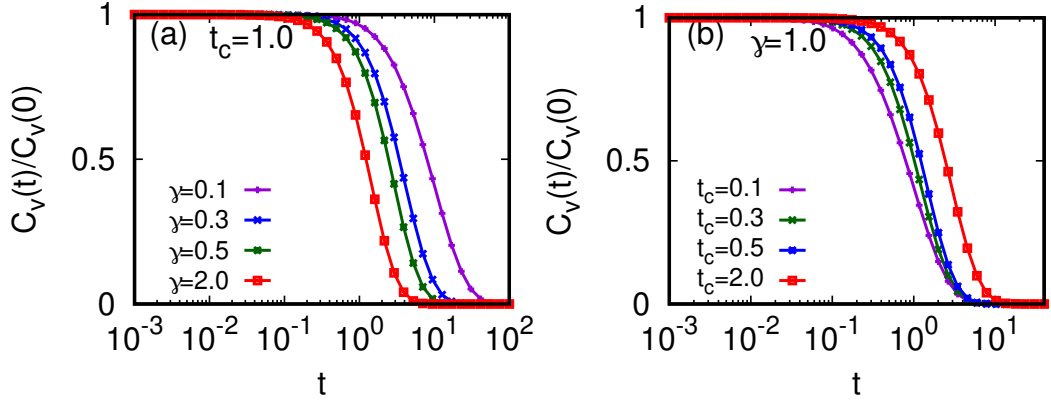}
    \caption{ Normalized $C_{v}(t)$[Eq.~\eqref{velocity correlation free}] as a function of $t$ for a free particle for different $\gamma$ values in (a) and for different values of $t_{c}$ in (b), respectively. The common parameters are $m=D=1$}
    \label{velocity correlation plot free}
    \end{figure*}
 The single particle trajectory of a freely moving self propelled particle in xy-plane is shown in Fig.~\ref{trajectoryfree} for different values of $\gamma$ as well as for different values of $t_{c}$. For a free particle, the motion is mainly governed by the inertial force and the self propulsion force and hence the time evolution of the dynamics is characterized by the complex interplay of the inertial time scale ($t_{m}=\frac{m}{\gamma}$) and the self propulsion time scale ($t_{c}$) of the dynamics. It is observed that for small $t_{c}$, when $\gamma$ is large, the particle shows a random trajectory. With either decrease in $\gamma$ or increase in $t_{c}$, the trajectory becomes smoother. Further, in Fig.~\ref{velocity correlation plot free}, we have plotted the normalized $C_{v}(t)$ as a function of $t$ for different values of $\gamma$ in (a) and for different values of $t_{c}$ in (b), respectively. Since the dynamics is governed by the inertial time and self propulsion time scale, $C_{v}(t)$ decays by the complex interplay of these two time scale. For any fixed $t_{c}$ or $\gamma$, it maintains with the same value in the lower time regimes before decaying to zero in the long time regimes.
 With increase in $\gamma$, $C_{v}(t)$ decays faster (see Fig.~6(a)) for a fixed $t_{c}$ and it's persistence with time decreases whereas with increase in $t_{c}$ it persists for longer time before decaying to zero as reflected from Fig.~6(b). It is always found to be positive for any chosen value of $\gamma$ or $t_{c}$. For very lower value of $t_{c}$, $C_{v}(t)$ is mainly controlled by the inertial time scale $t_{m}$. Hence in $t_{c} \rightarrow 0$ limit, $C_{v}(t)$ displays the same behaviour as in case of an inertial passive particle.
 
 Similarly the steady state correlation $C_{x}(t)$ corresponding to position of a free particle can be calculated as
    \begin{equation}
    C_{x}(t)=x_{0}^{2}+2x_{0}t_{m}v_{0}+\left(t_{m}v_{0}\right)^{2}+C_{x}^{'}(t),\label{position correlation free}
    \end{equation}
    where\\
    $ C_{x}^{'}(t)=\dfrac{D}{\gamma^{2}}\left[ t-t_{m}-t_{c}^{2}+\dfrac{t_{m}^{3}}{t_{c}^{2}-t_{m}^{2}}e^{\dfrac{-t}{t_{m}}}-\dfrac{t_{c}}{2}e^{\dfrac{-t}{t_{c}}}\left\lbrace1 +t_{m}'\right\rbrace  \right]$
   with  $t_{m}'=\dfrac{t_{m}}{t_{c}+t_{m}}$. It is worth to note that $C_{x}(t)$ depends on the initial velocity of the particle. Although $C_{x}^{'}$ is part of the correlation which depends on both $t_{m}$ and $t_{c}$ values, it is mainly controlled by the strength of the AOUP noise, $D$. Further using Kubo's theorem, we have calculated the diffusion coefficient as $D_{f}=\int_{0}^{\infty}\left\langle v(t)v(0)\right\rangle$
    =$\dfrac{D}{2\gamma^{2}}$. The diffusion coefficient depends only on the strength of noise as well as on the viscous drag of the medium. However, it is independent of the mass of the particle as reported in Ref.~\cite{nguyen2021active}.
    
    Next we are interested in the exact computation of MSD ($ \langle \Delta x^2(t)\rangle$) for both harmonically confined particle and free particle. The MSD for a harmonically confined particle ($\Delta x^2(t)\rangle_h$) is calculated as
    \begin{equation}
    \langle \Delta x^2(t)\rangle_h=\langle(x(t)-x_0)^2\rangle \label{MSD}
    \end{equation}
    \begin{widetext}
    \begin{equation}
     \begin{split}
 \langle \Delta  x^2(t)\rangle_h &= \frac{e^\frac{-t}{t_m}}{4 \omega_1^2t_m^2}\left[-2t_mx_0\omega_1e^\frac{t}{2t_m}+2t_mx_0\omega_1\cos(\omega_1t)+(2t_mv_0+x_0)\sin(\omega_1t)\right]^2 - \frac{D}{\gamma^2 t_m^2\omega_1^2}\left[\frac{\left(2-\frac{t_c}{t_m}\right)e^\frac{-t}{t_m}}{\left(\frac{t_c}{t_m}-2\right)^2+4t_c^2\omega_1^2}\right] \\
 & +\frac{D}{\gamma^2 t_m^2\omega_1^2}\left\{\frac{8\omega_1^2\left(1+\frac{t_c}{t_m}\right)}{\left[4\omega_1^2+\frac{1}{t_m^2}\right]\left[\left(2+\frac{t_c}{t_m}\right)^2+4t_c^2\omega_1^2\right]} + \frac{8t_c \omega_1\left[2t_c\omega_1\cos(\omega_1t)+\left(2+\frac{t_c}{t_m}\right)sin(\omega_1t)\right]e^{-t\left(\frac{1}{t_c}+\frac{1}{tm}\right)}}{\left[\left(-2+\frac{t_c}{t_m}\right) ^2+4t_c^2\omega_1^2\right]\left[\left(2+\frac{t_c}{t_m}\right) ^2+4t_c^2\omega_1^2\right]}\right\}\\
&+ \frac{D}{2\gamma^2\omega_1^2t_m^2}\frac{e^{\frac{-t}{t_m}}\left[\left(\frac{-4}{t_m}+\frac{2t_c}{t_m^2}-8t_c\omega_1^2\right)\cos(2\omega_1t)+8\omega_1\left(1-\frac{t_c}{t_m}\right)sin(2\omega_1t)\right]}{\left[4\omega_1^2+\frac{1}{t_m^2}\right]\left[\left(-2+\frac{t_c}{t_m}\right)^2+4t_c^2\omega_1^2\right]}\label{MSDharmonic}
\end{split}
\end{equation}
\newpage
\end{widetext}
By expanding equation.~\eqref{MSDharmonic} in the powers of $t$, it can be expressed as,
\begin{widetext}
\begin{equation}
\langle \Delta x^2(t)\rangle_h=v_0^2t^2-v_0\left(\frac{v_0}{t_m}+x_0\omega_0^2\right)t^3+\frac{1}{12}\left[\frac{3D}{2t_m^2t_c\gamma^2}+\frac{7v_0^2}{t_m^2}-4v_0^2\omega_0^2+\frac{10v_0x_0\omega_0^2}{t_m}+3x_{0}^2\omega_0^4\right]t^4+O\left(t^5\right)\label{MSDharmonic_series}
\end{equation}
\end{widetext}

In the long time limit (i.e., $t \rightarrow \infty$ limit), $\langle \Delta x^2(t)\rangle_h$ is found to be
\begin{equation}
\langle \Delta x^2(t)\rangle_h= x_0 ^2+\frac{D\left(1+\frac{t_c}{t_m}\right)}{4\gamma^2t_m^2\omega_0^2\left[\frac{1}{t_m}+\frac{t_c}{t_m}\left(t_c\omega_0^2+\frac{1}{t_m}\right)\right]}.
\label{MSDharmonicsteady}
\end{equation}
Similarly, the MSD for a free particle is obtained as
    \begin{widetext}
    \begin{equation}
    \begin{split}
\langle \Delta x^2(t)\rangle_f=\frac{D}{\gamma^2}\left[t-2t_{m}-t_c+t_ce^\frac{-t}{t_c}\right]+ v_0^2t_m^2 (e^{\frac{-t}{t_m}}-1)\\+\frac{Dt_ct_m}{2\gamma^2 (\text{tc}^2-\text{tm}^2)}\left[t_m e^\frac{-t}{tc}+3t_c e^\frac{-t}{t_m}+tc-t_ce^{-t(\frac{1}{t_c}+\frac{1}{t_m})}\right]-\frac{Dt_m^2}{4\gamma^2 (t_c^2-t_m^2)}\left[t_m+t_c e^{\frac{-t}{t_m}}+3t_me^{\frac{-t}{t_m}}\right]\label{MSDfree}
\end{split}
\end{equation}
\end{widetext}

\begin{figure}[!hb]
    \includegraphics[scale=0.46]{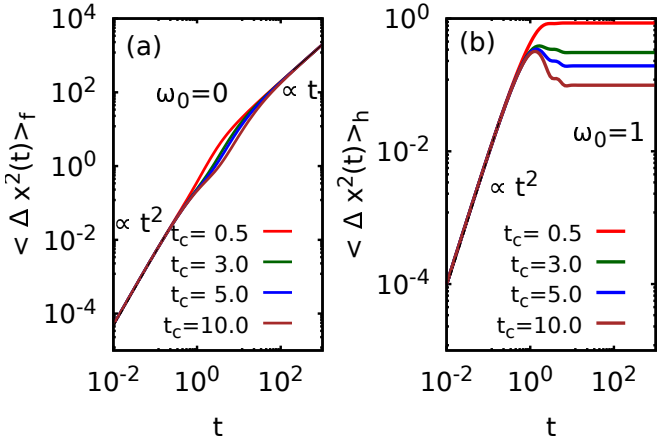}
\caption{MSD as a function of $t$ in (a) for a free particle [\text{Eq}.~\eqref{MSDfree}] and in (b) for a confined harmonic particle [\text{Eq}.~\eqref{MSDharmonic}], respectively for differnt values of $t_{c}$. The common parameters are $m=D=\gamma=1$}.
\label{mean square displacement}
\end{figure}
The power series expansion of \eqref{MSDfree} in the powers of $t$, can be written as
    \begin{equation}
\langle \Delta x^2(t)\rangle_f=t^2 v_0^2- \frac{t^3 v_0^2}{t_m}+t^4 \left(\frac{D}{8 m^2 t_c}+\frac{7v_0^2}{12t_m^{2}}\right)+O\left(t^5\right)\label{MSDfreeseries}
\end{equation}
In Fig.~\ref {mean square displacement}, we have shown the $\langle \Delta x^2(t)\rangle_f$ as a function of $t$ both for a free particle(Fig.~\ref {mean square displacement}(a)) and a confined harmonic particle (Fig.~\ref {mean square displacement}(b)) for different values of $t_{c}$. The calculation results for $\langle \Delta x^{2}(t) \rangle$ for both free and harmonically confined particle confirms that in the $t\rightarrow 0$  limit, $\langle \Delta x^{2}(t) \rangle$ is proportional to $t^{2}$ and hence the initial time regimes are ballistic in nature. The expression of MSD for a free particle in the long time limit (i.e., $t \rightarrow \infty$ limit) is obtained as $\langle \Delta x^2(t)\rangle_f= \frac{Dt}{\gamma^2}\label{MSDfreelongtime}$, i.e., proportional to $t$.  However, it approaches a constant value for a harmonically confined particle. Hence, the long time regimes for a free particle are diffusive where as for a confined harmonic particle, it is non diffusive in nature. Moreover, for a free particle the initial time regime as well as the long time regimes are not dependent on $t_{c}$, but there is a dependence of $t_{c}$ in the intermediate time regimes. The MSD for a harmonically confined particle in the time asymptotic limit depends on $t_{c}$ and gets suppressed with increase in $t_{c}$ value and finally approaches zero in $t_{c} \rightarrow \infty$ limit. This confirms that the particle gets more and more confined around the mean position of the well for longer persistence of activity in the medium.
    \section{Conclusion}
    In this work, we have studied the dynamics of a harmonically confined particle following the AOUP process. With the help of both numerical simulation and exact analytical calculations, we have explored the behaviour of particle trajectories, the steady state correlation functions, and mean square displacement both for a free particle and a harmonically confined particle. From the simulated trajectories, it is confirmed that the dynamics of a harmonic particle is mainly characterized by the inertial time, self propulsion time, and harmonic time scale. Similarly, for a free particle it is governed by the complex interplay of the inertial time and the self propulsion time. One of our important observations is that for a harmonic particle, in lower viscous medium, the particle makes circular trajectories due to strong influence of inertia. Further with increase in the self propulsion time of the dynamics, the circular trajectories get squeezed. This confirms that the particle gets a stronger confinement with the longer persistence of activity in the medium. The same conclusion is drawn from the behaviour of mean square displacement, which gets suppressed with increase in the self propulsion time. The steady state correlation functions shows oscillatory behaviour in the inertial regime and this oscillatory behaviour disappears either with viscosity of the medium or with self propulsion time of the dynamics. However, the velocity correlation shows a damped oscillatory behaviour in the over damped regime. In contrast to that for a free particle, the steady state velocity correlation is always found to be finite and positive before decaying to zero in the long time limit. When the inertial time and the self-propulsion time are of equal order, velocity of a free particle at two different times get decorrelated.  Further, it would be interesting to explore the relaxation behaviour of the system in various limits of the dynamics.
\section{Acknowledgment}
MS acknowledges the INSPIRE Faculty award (IFA 13 PH-66) by the Department of Science and Technology and the UGC Faculty recharge research grant for the financial support. We dedicate this work to our mentor, teacher and friend Arun Jayannavar, who passed away recently. \\
  \begin{appendix}
    \section{}
   For the confined harmonic particle, the dynamics [Eq.~\eqref{model1}] can be expressed in the form of a vector equation as\cite{caprini2021inertial},
    \begin{equation}
            \dot{n}=-A'\cdot n+\sqrt{2} \sigma' \cdot \eta' \label{appendix matrix form}
        \end{equation}
        where,
        \[
        n=\left( \begin{array}{c}
        x\\
v\\
\eta
 \end{array}\right),
        A'=\left( \begin{array}{ccc}
        0 & -1 & 0 \\
\frac{k}{m}&  \frac{\gamma}{m} & -\frac{\sqrt{D}}{m} \\
0 & 0 & \frac{1}{t_{c}}
\end{array} \right)
\text{and }
\sigma'=\left( \begin{array}{ccc}

0 & 0 & 0 \\
0&  0 & 0 \\
0 & 0 & \frac{1}{\sqrt{2}t_{c}}
\end{array} \right)
\] 

Using the correlation matrix method formalism, by introducing a correlation matrix $(C')$, Eq.~\eqref{appendix matrix form}can take a new from as, 
\begin{equation}
    A' \cdot C'+C'\cdot A'^{T}=2M',\label{appendix correlation matrix form}
\end{equation}
where $M^{'}=\sigma^{'}\cdot \sigma^{'T}$ is the diffusion matrix for confined particle and
\[
C^{'}=
\begin{pmatrix}
C^{'}_{xx} & C^{'}_{xv} & C^{'}_{x\eta}\\
C^{'}_{vx} & C^{'}_{vv} & C^{'}_{v\eta}\\
C^{'}_{\eta x} & C^{'}_{\eta v} & C^{'}_{\eta \eta}
\end{pmatrix}
.
\]
 Solving Eq.~\eqref{appendix correlation matrix form}, the elements of correlation matrix can be obtained as,
    $C'_{xx}=0$, $C'_{xv}=\frac{(m+\gamma t_{c})}{2\gamma^{2}(m+\gamma t_{c}+kt_{c}^{2})}=-C'_{vx}$, $C'_{x\eta}=C'_{\eta x}=\frac{\sqrt{D}t_{c}}{2(m+\gamma t_{c}+kt_{c}^{2})}$, $C'_{vv}=\frac{D}{2\gamma(m+\gamma t_{c}+kt_{c}^{2})}$, $C'_{v\eta}=C'_{\eta v}=\frac{\sqrt{D}}{2(m+\gamma t_{c}+kt_{c}^{2})}$  and  $ C'_{\eta \eta}=\frac{1}{2t_{c}} $.
The time dependent equation for velocity auto correlation function $\langle v(t) v(0) \rangle$ can be obtained from the Eq.~\eqref{model1} as,
 \begin{equation}
 \begin{split}
          \frac{d^{2}}{dt^{2}}\langle v(t)v(0)\rangle+\frac{\gamma}{m}\frac{d}{dt}\langle v(t)v(0)\rangle+\frac{k}{m}\langle v(t)v(0)\rangle=\\\frac{\sqrt{D}}{m}\langle\eta (t)v(0)\rangle.\label{derivative velocorre harmonic}
          \end{split}
      \end{equation}
     Solving the time dependent equation for $\langle \eta(t) v(0) \rangle$, one can obtain
      \begin{equation*}
          \langle\eta (t)v(0)\rangle=C'_{\eta v}\exp{-\frac{t}{t_{c}}}.
      \end{equation*}
     Now substituting the value of $C'_{\eta v} $ in the above equation, one can obtain
      \begin{equation}
      \langle\eta (t)v(0)\rangle=\frac{\sqrt{D}}{2(m+\gamma t_{c}+kt_{c}^{2})}\exp{-\frac{t}{t_{c}}}\label{etav correlation}
      \end{equation}
     Further substituting $\langle\eta (t)v(0)\rangle$ and solving Eq.~\eqref{derivative velocorre harmonic}, the correlation corresponds to velocity $C_{v}(t)=\langle v(t) v(0) \rangle$ can be obtained as
      \begin{equation}
C_{v}(t)=\dfrac{-f_{1}}{2m^{2}}e^{\dfrac{-t}{t_{c}}}+\frac{N}{4}e^{\dfrac{- t}{2t_{m}}}\left( f_{4} \cos(\omega_{1}t)+f_{5}\sin(\omega_{1}t)\right),\label{velocorrelation matrix method}
\end{equation}
      which is same as what we get using the Fourier transform method.
      \end{appendix}
%

\end{document}